\documentclass[12pt]{article}
\usepackage{graphicx}

\usepackage{dcolumn}
\usepackage{bm}
\usepackage{hyperref}
\usepackage{amssymb}
\usepackage{xcolor}
\usepackage{amsmath}

\newcommand\pubnumber{CIPANP2018-Doria}
\newcommand\pubdate{\today}

\def\kph{$^1$PRISMA Cluster of Excellence and Institut f\"ur Kernphysik,
Johannes Gutenberg-Universit\"at Mainz,
Johann-Joachim-Becher-Weg 45
D 55128 Mainz}
\def\him{$^2$Helmholtz Institute Mainz, Germany}

\def\Title#1{\begin{center} {\Large #1 } \end{center}}
\def\Author#1{\begin{center}{ \sc #1} \end{center}}
\def\Address#1{\begin{center}{ \it #1} \end{center}}

\newcommand\pubblock{\rightline{\begin{tabular}{l} \pubnumber\\
         \pubdate  \end{tabular}}}
\newenvironment{Abstract}{\begin{quotation}  }{\end{quotation}}
\newenvironment{Presented}{\begin{quotation} \begin{center} 
             PRESENTED AT\end{center}\bigskip 
      \begin{center}\begin{large}}{\end{large}\end{center} \end{quotation}}
\def\Acknowledgements{\bigskip  \bigskip \begin{center} \begin{large}
             \bf ACKNOWLEDGEMENTS \end{large}\end{center}}

\def\beq{\begin{equation}}
\def\eeq#1{\label{#1}\end{equation}}
\def\eeqn{\end{equation}}

\def\beqa{\begin{eqnarray}}
\def\eeqa#1{\label{#1}\end{eqnarray}}
\def\eeqan{\end{eqnarray}}

\let\bar=\overbar

\def\Dslash{\not{\hbox{\kern-4pt $D$}}}
\def\dslash{\not{\hbox{\kern-2pt $\del$}}}

\def\msb{{\bar{\ssstyle M \kern -1pt S}}}

\begin{document}
\begin{titlepage}
\pubblock

\vfill
\Title{Search for light dark matter with the MESA accelerator}
\vfill
\Author{Luca~Doria$^{1}$, Patrick~Achenbach$^{1,2}$, Mirco~Christmann$^{1,2}$,
  Achim~Denig$^{1,2}$, Pepe~G\"ulker$^{1}$, Harald~Merkel$^{1}$} 
\Address{\kph}
\Address{\him}

\vfill
\begin{Abstract}
At the Institute for Nuclear Physics of the Johannes Gutenberg University in Mainz,
the MESA facility is currently being constructed. At its core there is a new superconducting
energy-recovery linac which will provide intense electron beams for precision experiments
in subatomic physics. An important part of the MESA physics program consists in the search for a
"dark sector" which is a candidate explanation for the longstanding dark matter problem.
This report will highlight the MESA dark sector program, and in particular two experiments will be described.
The first one, MAGIX, is a two-spectrometer setup employing an internal gas-jet target
installed on a recirculation arc of MESA. The second one is a beam-dump experiment for directly detecting
dark matter particles. The experiments are in the R\&D phase and the current status and future prospects will be presented.
\end{Abstract}
\vfill
\begin{Presented}
CIPANP18, May 29 - June 3, 2018 Palm Springs, CA
\end{Presented}
\vfill
\end{titlepage}
\def\thefootnote{\fnsymbol{footnote}}
\setcounter{footnote}{0}

\section{Introduction}
Many compelling astrophysical observations support the existence of Dark Matter (DM)
through its gravitational effects.
In the last decades, DM detection eluded intensive and diverse searches at
particle colliders such as for instance LHC \cite{dmLHC},
fixed-target experiments, B-factories, and underground
direct detection experiments \cite{dmdirect}.
A class of models describes DM as a relic from an epoch where DM was in thermodynamic
equilibrium with SM particles in the early Universe. DM abundance was set when its
annihilation rate in SM particles became smaller than the expansion rate of the universe,
a process known as {\em freeze-out}. Although this mechanism is very compelling, it allows for
a very broad range of DM masses and interaction cross-sections (keV/c$^2$-TeV/c$^2$).
The $>$GeV/c$^2$ mass range is commonly connected to the electroweak scale and to models
trying to address the hierarchy problem (most notably supersymmetric models).
This range can be effectively tackled by high-energy colliders and direct detection experiments.
In the $<$GeV/c$^2$ range these techniques become less effective, with the lowest thresholds
for DM-nucleon scattering obtained with cryogenic silicon and germanium detectors \cite{scdms}.
Experiments at high-energy colliders are not optimized for the low mass range: a typical DM
signature would be the detection of missing mass, which is too small to be detected
given the backgrounds present.
Direct detection experiments are mostly based on the nuclear recoil signal from galactic DM scattering,
and for low masses it becomes too small to be detected, although
more recently, notable progress has also been made on DM-electron scattering \cite{xenon,ecdms}.
For the $<$GeV/c$^2$ Light Dark Matter (LDM) range, for retaining
its thermal origin we have to postulate the existence of additional interactions.
Considering only electroweak-scale cross-sections, the annihilation rate would not
be sufficient, leading to DM overproduction.
Interesting LDM models are based on the idea that DM particles belong to a {\em dark sector}
interacting with the SM via one (or more) mediator particle(s).
Recently, dark sector models where the mediator decays visibly to SM particles were 
the subject of intense experimental exploration \cite{dmreport}.
The case where the mediator
decays invisibly in other dark sector particles (for example in DM particles) is
less explored, leaving ample room for new experiments.
Accelerator experiments are well-suited for investigating LDM, since the high-momentum
particle production allows to circumvent the low-threshold problem.
In this work, we discuss two experiments aimed at the detection of LDM,
with special focus on a newly proposed beam-dump experiment.




\section{Theoretical Model}
Dark sector models can be distinguished from the type of the mediator particle
(the {\em portals}) and the type of DM particle.
In general, the dark sector can contain more mediators and particles.
Here we focus on a simple model which still captures
the essence of dark sector physics.
The model is comprised by a massive vector mediator particle (a "dark photon")
which can be thought as a gauge boson resulting from a spontaneously broken $U(1)_D$
symmetry. The model's lagrangian is
\begin{equation}
  \begin{aligned}
  \mathcal{L}_{A'}&=-\frac{1}{4}F'_{\mu\nu}F'^{\mu\nu}+\frac{\epsilon_Y}{2}F'_{\mu\nu}B^{\mu\nu}\\
  &+\frac{m^2_{A'}}{2}A'_{\mu}A'^{\mu} + g_DA'_{\mu}J_{\chi}^{\mu}+g_YB_{\mu}J^{\mu}_Y \quad,
  \end{aligned}
\end{equation}
where $F'_{\mu\nu} = \partial_{\mu}A'_{\nu}- \partial_{\nu}A'_{\mu}$ and
$B_{\mu\nu} = \partial_{\mu}B_{\nu}- \partial_{\nu}B_{\mu}$ are
respectively the dark photon and the hypercharge fields, $g_D$ is the dark gauge coupling,
and $J_{\chi}^{\mu}$ and $J^{\mu}_Y$ the DM and hypercharge currents, respectively.
After electroweak symmetry breaking, the dark photon mixes with the SM photon and the Z boson
\begin{equation}
  \frac{\epsilon_Y}{2}F'_{\mu\nu}B^{\mu\nu} \rightarrow \frac{\epsilon}{2}F'_{\mu\nu}F^{\mu\nu} +
  \frac{\epsilon_Z}{2}F'_{\mu\nu}Z^{\mu\nu} \quad,
\end{equation}
where $\epsilon=\epsilon_Y / \cos\theta_W$, $\epsilon_Z=\epsilon_Y/\sin\theta_W$, and
$\theta_W$ is the weak mixing angle. After diagonalization, the coupling of the dark
photon with DM and the SM photon is
\begin{equation}
  g_DA'_{\mu}J_{\chi}^{\mu}+g_YB_{\mu}J^{\mu}_Y \rightarrow A'_{\mu}(g_DJ_{\chi}^{\mu}+\epsilon e J_{EM}^{\mu}) \quad,
\end{equation}
where $J_{EM}^{\mu}$ is the SM electromagnetic current. The DM current $J_{\chi}^{\mu}$ content depends
on the exact nature of the DM particle. The coupling of the dark photon to SM particles happens
via the "millicharge" $\epsilon e$, while the dark fine structure constant $\alpha_D=\sqrt{4\pi g_D}$
describes the coupling with DM.\\
It is useful to define the dimensionless combination of the model parameters
\begin{equation}
  y = \epsilon^2\alpha_D\left( \frac{m_{\chi}}{m_{A'}} \right)^4 \quad,
\end{equation}
which is proportional to the thermally averaged DM annihilation rate. 

\begin{figure*}
  \begin{center}
    \includegraphics[scale=0.3]{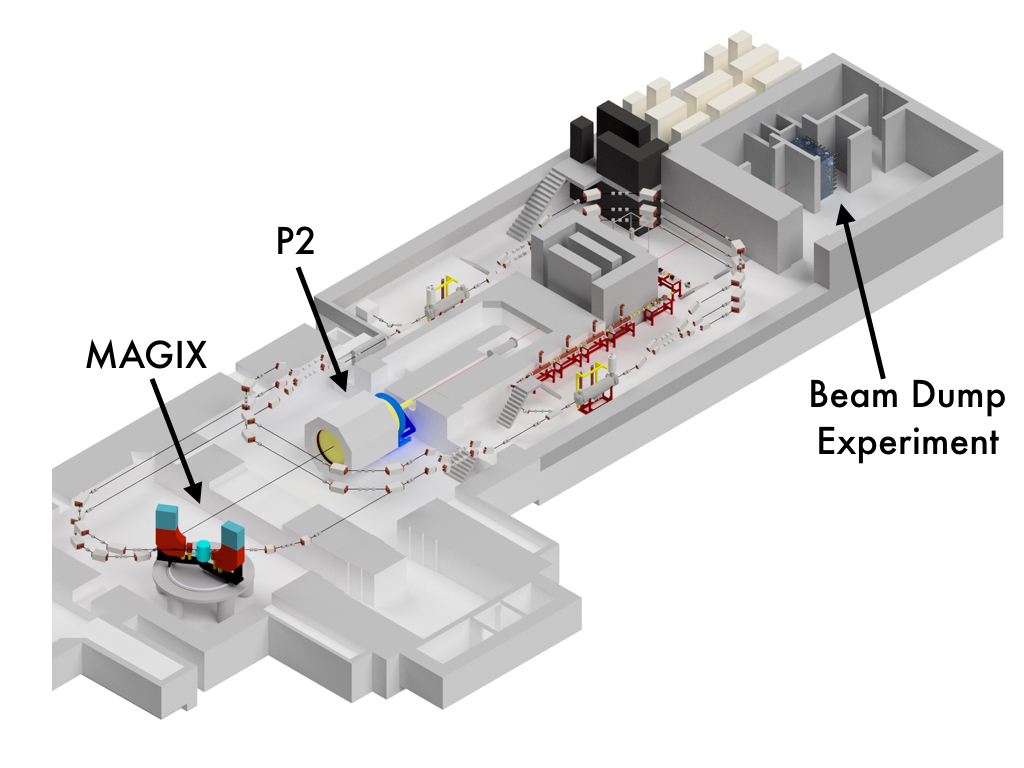}
  \end{center}
  \caption{The MESA (Mainz Energy-Recovery Superconducting Accelerator) complex with the three foreseen experiments.}
  \label{fig:MESA}
\end{figure*}

\section{The MESA accelerator}
The Institute for Nuclear Physics at Mainz University is building a new
CW multi-turn energy recovery linac for precision particle physics experiments
with a beam energy range of 100-200 MeV.
MESA will operate in two modes: energy recovery mode (ERM) and external beam mode (XBM).
In ERM, the accelerator will provide a beam current of up to 1~mA at 105 MeV for the MAGIX internal
target experiment with multi-turn energy recovery capability.
In XBM, a polarized beam of 150 $\mu A$ will be provided to the P2 experiment \cite{P2}. In this
mode, the initial design energy is up to 155 MeV.
The linac will provide an energy gain of 50 MeV/pass by using four ELBE-like 9-cell cavities \cite{cavities}
installed in two cryomodules. 

\begin{figure}[t!]
  \centering
\includegraphics[width=0.7\textwidth]{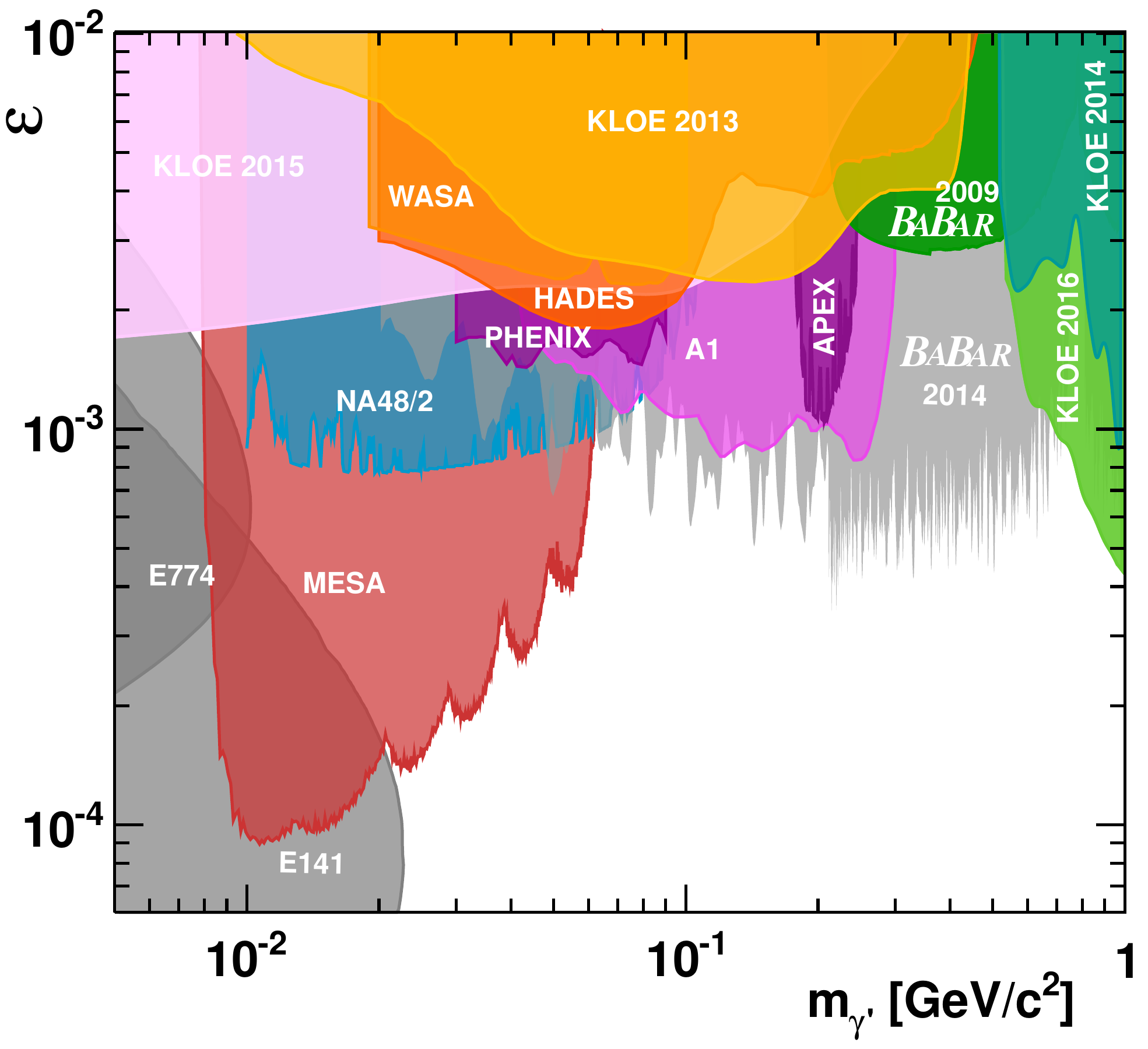}
\caption{MAGIX projected exclusion limits for the $\gamma'\rightarrow e^+e^-$ decay.}
\label{fig:magix}
\end{figure}

\section{The MAGIX Experiment}
MAGIX is a versatile experiment taking advantage from the unique combination of a gas-jet target and
the MESA CW beam in energy recovery mode.
The experiment is based on two magnetic spectrometers with resolution $\delta p / p \sim 10^{-4}$
and low-material budget focal plane detectors.
MAGIX will allow precision measurements on a variety of fields
ranging from hadron physics to nuclear astrophysics and dark sector searches.
The dark photon can be produced through a mechanism similar to bremsstrahlung on a heavy
nuclear target Z via the reaction $e^-Z\rightarrow e^-Z\gamma^{\prime}$. If the dark photon decays
into SM particles, e.g. $\gamma^{\prime}\rightarrow e^+e^-$, the electron/positron final state 
can be detected in coincidence in the two spectrometers. A peak-search on the QED background can thus be
performed \cite{A1}. If the dark photon decays invisibly
(e.g. into light dark matter particles $\gamma^{\prime}\rightarrow \chi\bar{\chi}$), this will require
the measurement of the recoil target nucleus. A peak-search on the reconstructed missing
mass $m_{\gamma^{\prime}}^2=(p_{beam}-p_{Z}-p_{e'})^2$ will be performed. 
Fig.~\ref{fig:magix} shows the projected sensitivity of MAGIX to the dark photon's visible decays.

\section{The Beam-Dump Experiment}
In a beam-dump experiment, the dark photon can be produced
radiatively by an impinging beam on a heavy nucleus Z through a process analogous to bremsstrahlung \cite{Bjorken:2009}.
The beam can be made of protons or electrons and
in the proton case the dark photon can be produced also in meson decays.
Here we focus on the electron beam case: $eZ\rightarrow eZ\gamma'$
and we assume that the dark photon decays into pairs of DM particles ($\gamma' \rightarrow \chi\bar{\chi}$).
Depending on the model, $\chi$ and $\bar\chi$ can be a particle/antiparticle couple or two
different particles $\chi_1$ and $\chi_2$ (inelastic DM \cite{iDM}).\\
After production, DM particles can be detected within a shielded detector downstream of the
beam-dump, via $e\chi \rightarrow e\chi$ and $p\chi \rightarrow p\chi$ scattering, where $p$
is a proton.\\
The dark photon production yield scales as $Y_{\gamma'} \sim \alpha^3\epsilon^2 / m^2_{A'}$
while the DM yield $Y_{\chi}$ in the detector is proportional to $\epsilon^2$, giving
a total number of detected DM particles scaling as $Y_{\gamma'}\cdot Y_{\chi} \sim \epsilon^4$. While the detection
yield does not have a favorable scaling, a beam-dump experiment has distinct advantages.
The large number of electrons on target (EOT) deliverable in a reasonable amount
of time by modern CW electron accelerators can compensate for the small yield
and reach high sensitivity.
Another advantage is provided by the boost at which DM particles are produced, allowing
an improved reach at low masses.
Moreover, such experiments are unique since they can probe at the same time both the dark
photon production and the DM interaction.
At MESA, a radiation-shielded area is available 23~m downstream of the beam-dump of the P2 experiment,
allowing for the installation of a detector for LDM searches (see Fig.~\ref{fig:MESA}).
A similar experiment is currently being discussed at Jefferson Lab \cite{bdx}.

\section{Simulation of the Beam-dump Experiment}
For assessing the sensitivity of a beam-dump experiment at MESA, a full simulation study was performed.
The {\tt Geant4} \cite{GEANT} simulation implemented the geometry of the experimental halls, the
relevant details of the P2 experiment (liquid hydrogen target and magnetic field),
the (mainly aluminum) beam-dump, and the detector.
{\tt MadGraph4} \cite{MG4} was used for generating the dark photon bremsstrahlung process.
The number of produced $\chi\bar{\chi}$ pairs per EOT  is
\begin{equation}
  N_{\chi\bar{\chi}}=\frac{N_A\rho_{BD} X_0}{A}\int_0^{T_{BD}}dt\int_{E_{min}}^{E_B}dE\sigma(E)\frac{dN}{dE}(t)\quad,
  \label{Nchi1}
\end{equation}
where $\rho_{BD}$ and $X_0$ are the beam-dump density and radiation length, respectively.
$T_{BD}$ is the beam-dump length in radiation length units, $\sigma(E)$ the $eN\rightarrow eN\gamma'\rightarrow eN\chi\bar{\chi}$
cross-section, $E_B$ the beam energy, and $E_{min}$ the minimum threshold energy.
Defining the {\em differential track length} (DTL)
\begin{equation}
  \langle \frac{dN}{dE} \rangle=  \int_0^{T_{BD}}dt \frac{dN}{dE}(t) \quad,
  \label{DTL}
\end{equation}
Eq.~\ref{Nchi1} becomes
\begin{equation}
  N_{\chi\bar{\chi}}=\frac{N_A\rho_{BD} X_0}{A}\int_{E_{min}}^{E_B}dE\sigma(E) \langle \frac{dN}{dE} \rangle \quad.
  \label{Nchi2}
\end{equation}
A common simplifying approximation, referred as "single-radiation length approximation"
assumes $\langle \frac{dN}{dE} \rangle= \delta (E-E_B)$, which neglects all the showering
effects in the beam-dump. These effects are quite important in assessing the experimental sensitivity,
since they can substantially reduce the dark photon production.
In the present simulation, the DTL was reconstructed using {\tt Geant4}:
the beam dump was divided in thin slices along the beam direction and at every
slice the $e^+$ and $e^-$ flux was calculated.
The sum of the fluxes allowed a numerical evaluation of the integral in Eq.~\ref{DTL} which resulted
in an energy spectrum characterizing the beam dump and the incident beam.

Fixing the parameters of the theoretical model ($m_{\gamma'}$, $m_{\chi}$, $\epsilon$, $\alpha_D$),
$\sigma(E)$ and the final state four-vectors were calculated with {\tt MadGraph4} for each bin of the DTL spectrum.
The results for each energy bin were summed with the respective weight given by the DTL spectrum.

The final state four-vectors for the $\chi /\bar{\chi}$ particles were re-introduced in the {\tt Geant4} simulation
where they were tracked through the various materials up to the detector location.
The $\chi /\bar{\chi}$ interaction with electrons or protons in the detectors was calculated
with a custom code embedded into {\tt Geant4} implementing the $e\chi$ and $p\chi$
cross-sections at first order with the exchange of a dark photon.

The total number of detected DM particles (Eq.~\ref{Nchi2}) was calculated as
\begin{equation}
  \begin{aligned}
    N_{\chi\bar{\chi}}  = EOT \times N_{D} \times N_{DET} \times N_{BD}\\
    \times X_0 \times \frac{\sigma_{MG}}{N_{SIM}} \times \sum_{i=0}^{i=N_D}L_i\sigma_i \quad,
  \end{aligned}
  \label{Nchifinal}
\end{equation}
where EOT is the number of electrons on target, $N_{D}$ is the number of $\chi /\bar{\chi}$ within the detector acceptance, $X_0$
the beam-dump radiation length, $L_i$ the track length in the detector of the $i-th$ DM particle track, $\sigma_i$
the $e\chi\rightarrow e\chi$ or $p\chi\rightarrow p\chi$ cross section of the $i-th$ DM particle track, $\sigma_{MG}$ the
$eA\rightarrow eA\gamma'\rightarrow \chi\bar{\chi}$ cross-section calculated with {\tt MadGraph4}, and $N_{SIM}$ the total
number of simulated events.
For a detector with a combination of materials with average atomic number Z, mass number A, and density $\rho_D$,
the total number of scattering centers (number of electrons or protons) is $N_{DET}=Z\rho_DN_A/A$,
where $N_A$ is the Avogadro number. With the same notation, the number of nuclei in the beam dump is $N_{BD}=\rho_{BD}N_A/A$.

\begin{figure}[!t]
\centering
\includegraphics[width=0.8\textwidth]{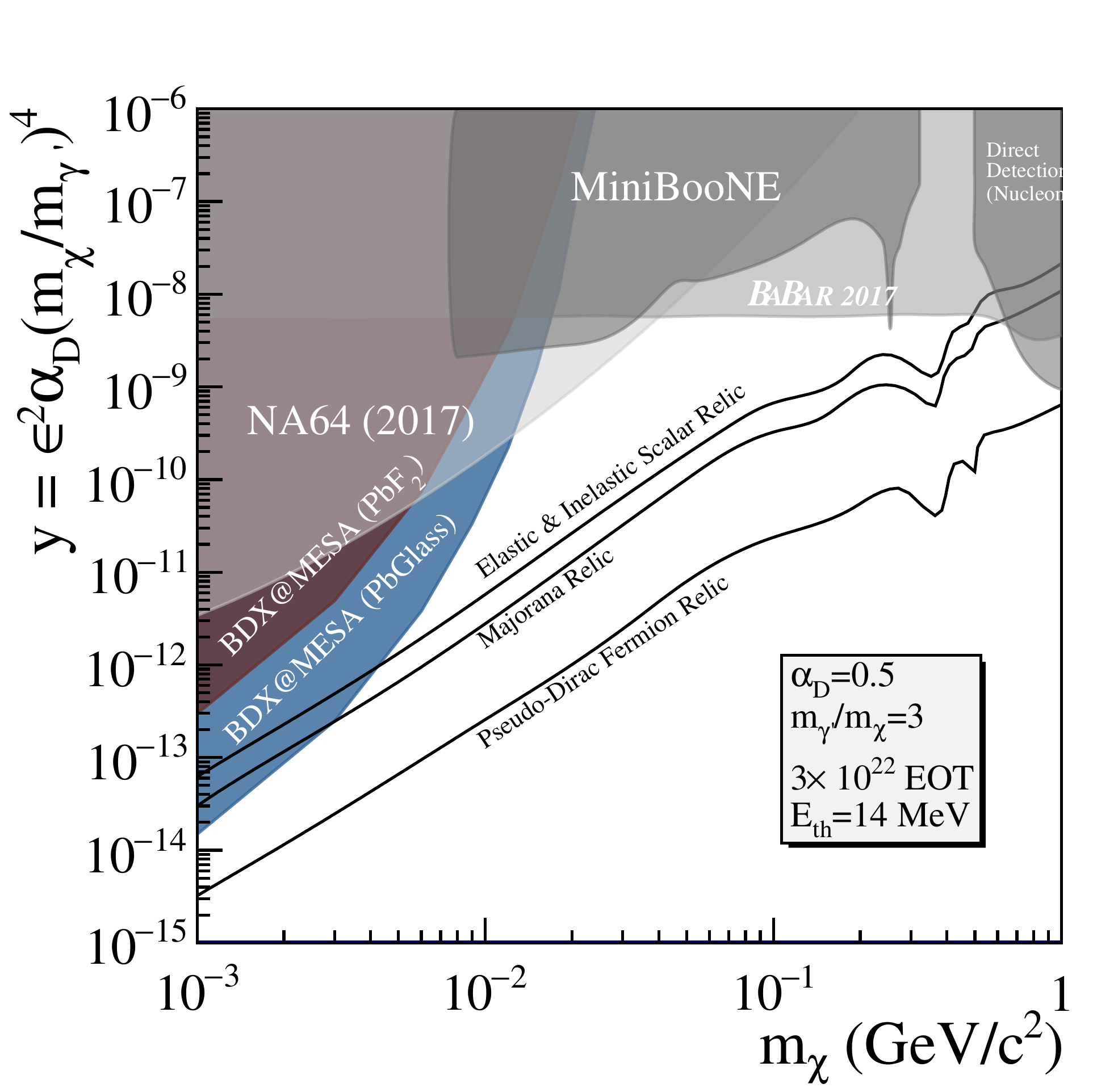}
\caption{Upper limits on the thermal target variable $y$ as a function of the dark matter particle mass $m_{\chi}$ for
  the two foreseen phases of the MESA beam-dump experiment.
The simulated limits are compared with the existing limits \cite{na64-1,na64-2,babar,minib,cresst}.
The black thermal relic target lines mark for a given $y-m_{\chi}$ combination and DM type
the limit to the annihilation cross-section for reproducing today's DM relic density.}
\label{upperlimits}
\end{figure}

\section{Simulated Beam and Detector}
The simulation employed a monoenergetic 155~MeV pencil electron beam hitting
the 60-cm long P2 hydrogen target. The $\sim$0.6~T solenoidal magnetic field of P2 was simulated and its
effect resulted in focusing the particles emerging from the target on the beam-dump. The beam-dump was
made of aluminum and cooling water and in front of it there was a 2.5~mm tungsten plate which thickness
was optimized for enhancing dark photon production over normal bremsstrahlung.\\
The current plan for the construction of the experiment is divided in two phases. Phase-1 will employ
already available PbF$_2$ crystals for building a $(1\times 1\times 0.13)$m$^3$ detector. The Phase-2 detector
will be constructed with Pb-Glass crystals with a $(2.7\times 2.7\times 1.5)$m$^3$ volume.
The advantage of Cherenkov crystals is their speed and relatively low sensitivity to background neutrons.
The simulation considered only DM-electron recoils, for which a significant amount of energy
was deposited in the calorimeter. DM-nuclei recoils mostly happened below the detector threshold of 14~MeV,
which was determined by a combination of electron beam tests and simulation.
The EOT used in Eq.~\ref{Nchifinal} was $3\times 10^{22}$ which corresponding to 10000~h of
measurement with a beam current of 150~$\mu$A ($\sim$5400~C total deposited charge).

\section{Results}
The results of the simulation for the two phases are summarized in Fig.~\ref{upperlimits} together with
the existing limits \cite{na64-1,na64-2,babar,minib,cresst}.
The 3-$\sigma$ exclusion limits on the thermal target variable $y$ are calculated
as a function of the dark matter mass $m_{\chi}$ conservatively assuming $m_{\gamma^{\prime}}=3m_{\chi}$,
$\alpha_D=0.5$ and a detector threshold $E_{min}$=14~MeV.
In this study, we did not assume the presence of backgrounds: this will be the subject of future investigations.
Particularly important are cosmics, environment, and detector backgrounds. The experiment requires
an efficient cosmics veto system and the detector threshold should be high enough in order
to exclude natural radioactivity backgrounds. The experiment will work below the pion production
threshold, thus neutrino or muon backgrounds from the beam-dump will be absent.
Phase-2 limits are approximately two orders of magnitude stronger than Phase-1, reflecting the
corresponding difference in detector volume, while the density is comparable.

\section{Summary}
The new MESA accelerator at the Institute for Nuclear Physics of the Johannes Gutenberg University in Mainz
will allow new exciting opportunities in precision tests of the Standard Model, nuclear physics,
as well as in new physics searches connected to the long-standing dark matter problem.
The MAGIX two-spectrometer setup, exploiting the high luminosity provided by MESA in combination
with a gas-jet target, will be able to search for visible and invisible decays of the dark photon in a
new mass range.
The installation of a beam-dump experiment represents an unique opportunity to expand the MESA research program
with a competitive experiment working parasitically to the other ones, taking
advantage of the world-class EOTs delivered by the new accelerator.
Moreover, beam-dump experiments have the advantage of being able to investigate at the
same time the production of the dark photon, its decay, and the DM interaction.

\Acknowledgements
We are grateful for the discussions and help with the {\tt MadGraph} simulation
to M.~Battaglieri and A.~Celentano.
This work is supported by the Deutsche Forschungsgemeinschaft with the
Collaborative Research Center 1044, the PRISMA Cluster of Excellence
"Precision  Physics,  Fundamental  Interactions and Structure of Matter",
and the Helmholtz-Institut Mainz.

\end{document}